\documentclass[conference]{IEEEtran}
\IEEEoverridecommandlockouts
\usepackage{cite}
\usepackage{amsmath,amssymb,amsfonts}
\usepackage{algorithmic}
\usepackage{newtxtext,newtxmath}
\usepackage{graphicx}
\usepackage{textcomp}
\usepackage{balance}
\usepackage{xcolor}
\def\BibTeX{{\rm B\kern-.05em{\sc i\kern-.025em b}\kern-.08em
    T\kern-.1667em\lower.7ex\hbox{E}\kern-.125emX}}
\begin{document}

\title{Requirements Development and Formalization for Reliable Code Generation: A Multi-Agent Vision
\thanks{*Weisong Sun is the corresponding author. This research is supported by the National Natural Science Foundation of China (62192734, 62192731, 62372347), the Fundamental Research Funds for the Central Universities (ZYTS25076), and the National Research Foundation, Singapore, and DSO National Laboratories under the AI Singapore Programme (AISG Award No: AISG2-GC-2023-008).}
}

\author{\IEEEauthorblockN{Xu Lu}
\IEEEauthorblockA{\textit{ICTT and ISN Lab} \\
\textit{Xidian University}\\
Xi'an, China \\
xlu@xidian.edu.cn}
\and
\IEEEauthorblockN{Weisong Sun*}
\IEEEauthorblockA{
\textit{Nanyang Technological University}\\
Singapore \\
weisong.sun@ntu.edu.sg}
\and
\IEEEauthorblockN{Yiran Zhang}
\IEEEauthorblockA{
\textit{Nanyang Technological University}\\
Singapore \\
yiran002@e.ntu.edu.sg}
\and
\IEEEauthorblockN{Ming Hu}
\IEEEauthorblockA{\textit{Singapore Management University}\\
Singapore \\
ecnu\_hm@163.com}
\and
\IEEEauthorblockN{Cong Tian}
\IEEEauthorblockA{\textit{ICTT and ISN Lab} \\
\textit{Xidian University}\\
Xi'an, China \\
ctian@mail.xidian.edu.cn}
\and
\IEEEauthorblockN{Zhi Jin}
\IEEEauthorblockA{\textit{Wuhan University}\\
Wuhan, China \\
zhijin@whu.edu.cn}
\and
\IEEEauthorblockN{Yang Liu}
\IEEEauthorblockA{
\textit{Nanyang Technological University}\\
Singapore \\
yangliu@ntu.edu.sg}
}

\newcommand{\ours}[1]{\textsc{ReDeFo}}

\maketitle

\begin{abstract}
Automated code generation has long been considered the holy grail of software engineering. The emergence of Large Language Models (LLMs) has catalyzed a revolutionary breakthrough in this area. However, existing methods that only rely on LLMs remain inadequate in the quality of generated code, offering no guarantees of satisfying practical requirements. 
    They lack a systematic strategy for requirements development and modeling. Recently, LLM-based agents typically possess powerful abilities and play an essential role in facilitating the alignment of LLM outputs with user requirements. In this paper, we envision the first multi-agent framework for reliable code generation based on \textsc{re}quirements \textsc{de}velopment and \textsc{fo}rmalization, named \ours{}. This framework incorporates three agents, highlighting their augmentation with knowledge and techniques of formal methods, into the requirements-to-code generation pipeline to strengthen quality assurance. The core of \ours{} is the use of formal specifications to bridge the gap between potentially ambiguous natural language requirements and precise executable code. \ours{} enables rigorous reasoning about correctness, uncovering hidden bugs, and enforcing critical properties throughout the development process. In general, our framework aims to take a promising step toward realizing the long-standing vision of reliable, auto-generated software.
\end{abstract}


\section{Introduction}

Automated Code Generation (ACG)~\cite{church1962logic} has long been regarded as an elusive yet highly desirable goal in software engineering, promising to dramatically improve productivity and reduce development effort. The recent rise of Large Language Models (LLMs) has cast a new light on this vision, offering unprecedented capabilities in understanding, reasoning, and synthesizing code that align with human intent~\cite{bi2024deepseek,achiam2023gpt}. LLM-based or LLM-agent-based ACG techniques accelerate software development by an end-to-end transformation from high-level specifications, which are typically expressed in Natural Language Requirements (NLRs) to executable code~\cite{jiang2024survey,hong2023metagpt,qian2023chatdev,wei2024requirements}. Its potential lies in boosting development efficiency, reducing human error, and enabling rapid prototyping. However, despite these advantages, the quality and correctness of generated code often remain uncertain. The primary challenge is the lack of guarantees about the consistency between high-level specifications and code, which can lead to unreliable software and elevated maintenance costs~\cite{liu2023your}. Therefore, the need for a sound transition from ambiguous NLRs to precise executable code is highlighted, with a systematic strategy for requirements development and modeling being especially essential.

Formal methods provide a rigorous approach to establishing software correctness by mathematically proving that a program adheres to its formal specifications \cite{clarke1996formal}. Applications of formal methods in the broader domain of program verification have achieved notable success, including detecting subtle bugs, enforcing safety and security guarantees, and supporting high-assurance software development \cite{woodcock2009formal}. Naturally, this leads us to consider whether ACG can be integrated with formal methods, as such integration could revolutionize the way we produce software by ensuring that the generated code is not only syntactically correct but also semantically faithful to the given requirements.

In this paper, we propose a multi-agent framework, \textsc{re}quirements \textsc{de}velopment and \textsc{fo}rmalization (\ours{}), to address this challenge. The framework begins with NLRs and proceeds through formal specification \cite{nissanke2012formal} to verified code generation. It leverages distinct types of agents together that work collaboratively to bridge the gap between informal human intent and provably correct software artifacts. To the best of our knowledge, this is the first multi-agent framework for ACG with formal correctness guarantees.

\section{\ours{} Framework}

\begin{figure*}[!t]
    \centering
    \includegraphics[width=1\linewidth]{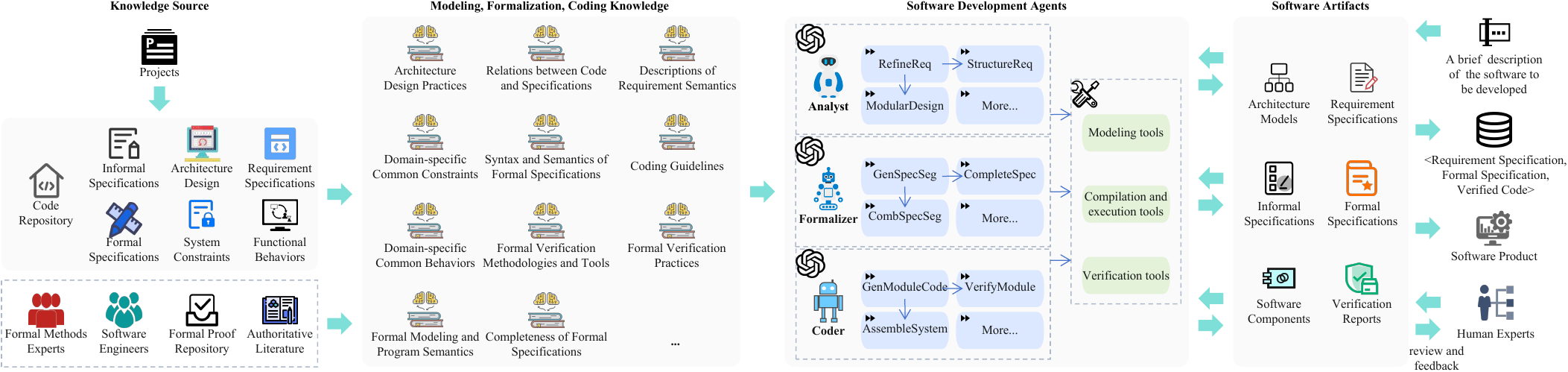}
    \caption{Overview of the proposed framework \ours{}}
    \label{fig: overview}
\end{figure*}

The overview of \ours{} is illustrated in Figure~\ref{fig: overview}, which depicts a multi-agent pipeline designed to transform NLRs to verified code. The overall process involves three functionally specialized agents, i.e., Analyst, Formalizer, and Coder who are responsible for completing the requirements development, formalization, and code generation tasks, respectively.

Upon receiving user-provided NLRs, the \textbf{Analyst} autonomously analyzes the text to refine each functionality into more detailed sub-requirements~\cite{2025-KGMAF, ataei2024elicitron}. Based on this understanding, it further extracts or infers system constraints and functional behaviors, identifies architectural components, and constructs an initial architecture model. The aforementioned results are then organized and summarized into informal, yet structured specifications. Further, the \textbf{Formalizer} is responsible for translating the informal specifications into formal specifications using domain-specific logic, formal semantics, and modeling guidelines. Finally, the \textbf{Coder} uses both the refined NLRs and formal specifications to generate code modules that meet the given requirements. The agent synthesizes executable code from the formal models, while retaining alignment with the original informal descriptions. Formal verification is then performed on the architecture model and generated code to guarantee correctness with respect to the specifications.

When facing ambiguous requirements, conflicting constraints, or uncertain decisions, agents can pose clarification questions or request confirmations from domain experts. \ours{} advocates human-in-the-loop interaction by incorporating expert judgment at critical points and facilitating more active expert participation beyond post-hoc review. Based on feedback produced throughout the pipeline, software engineers and formal methods experts can either review intermediate artifacts or respond to clarification questions and confirmation requests posed by the agents. The feedback may comprise verification failures, unmet constraints, or inconsistencies between artifacts, etc. Through a modular and collaborative workflow, the framework produces a comprehensive set of artifacts, including requirement specifications, architecture model, informal and formal specifications, verified architecture model/software components, and verification reports, thus culminating in a trustworthy software product. Moreover, the requirements and formal specifications, along with the verified code, constitute a high-quality corpus of aligned data.

\subsection{Agents in \ours{}}

\subsubsection{Analyst} 

Acting as a pivotal agent in \ours{}, the Analyst is responsible for interpreting and structuring the NLRs. To fulfill its role effectively, the Analyst performs the following actions: (1) Refinement and slicing of requirements, where each function is analyzed in detail to uncover implicit constraints or behaviors and express them as self-contained, formalization-ready sub-requirements. (2) Elaboration of vague or underspecified requirements, refining ambiguous or high-level descriptions into concrete, actionable statements. (3) Identification of system constraints or behaviors that are directly formalizable. They are critical properties that can be expressed in precise terms and mapped relatively easily into formal representations. (4) Functional decomposition and architectural design, whereby the overall system is broken down into modular components that reflect logical or domain-specific divisions.

The above actions are supported by a range of necessary domain knowledge and engineering expertise: (1) Basic understanding of formal specification languages and their capabilities, enabling the identification of formalizable content within informal text. (2) Methods for extracting semantic meaning from NLRs, including structured techniques in requirements engineering \cite{laplante2022requirements} such as use case analysis, goal modeling, and scenario-based modeling. (3) Skills in drafting requirements that closely mirror formal semantics, thereby easing the translation to formal specifications in the next phase. (4) Expertise in system architecture and modular design, allowing the agent to produce a viable and maintainable component layout.

\subsubsection{Formalizer}

The Formalizer undertakes the task of translating informal requirements into precise, unambiguous formal specifications. It engages in several key actions to accomplish the task: (1) Selection of suitable formalisms and generation of formal specification segments from informal requirement slices \cite{hahn2022formal, aaai2023fc, cao2025informal, wu2022autoformalization}. (2) Composition of specification segments by merging similar or overlapping statements that describe the identical or logically entailed underlying semantics, thus preserving coherence and reducing redundancy. (3) Completion of specifications, which serves to infer and explicitly describe implicit semantics not fully captured in the informal requirements. (4) Consistency assessment, whereby the agent evaluates the logical soundness and completeness of the specifications to detect contradictions, gaps, or ambiguities.

The actions rely on a solid foundation of domain-specific formal knowledge: (1) Proficiency in formal specification syntax and semantics, including understanding of popular formalisms used in model checking~\cite{baier2008principles}, theorem proving~\cite{loveland2016automated}, static analysis~\cite{rival2020introduction}, etc. (2) Methodologies for formally describing system properties cover both functional and non-functional aspects, such as invariants~\cite{donaldson2011software}, pre/post-conditions~\cite{hoare1969axiomatic}, and temporal logic assertions~\cite{rescher2012temporal}. (3) Strategies for evaluating soundness and completeness of specifications \cite{DBLP:journals/tse/MenghiBVSR25}, e.g., structural checks, behavioral coverage analysis, and traceability/consistency verification. (4) Semantic interpretation skills, supporting the detection of misalignments between the informal intent and its formalized representation.

\subsubsection{Coder}
\label{subsubsec:coder}
The Coder is the final component in \ours{}, tasked with programming the specifications into verified, executable code. The Coder organizes a series of coordinated actions: (1) Module code generation, where the agent synthesizes executable code based on the requirements and formal specifications for each module~\cite{murphy2024combining}. (2) Code \& model verification, whereby the generated modules and models are checked against the formal specifications using appropriate formal verification techniques. (3) System assembly, in which individual modules are integrated into a complete software.

The Coder requires profound technical and formal understanding, including: (1) Professional practices in formal verification methodologies as well as their respective strengths, limitations, and applicable contexts. For example, using model checking for state-based systems with bounded complexity, or theorem proving for systems requiring high expressiveness and rigor. (2) Mapping strategies from formal specification to code, possibly including language-specific implementation patterns and automated code synthesis techniques~\cite{bloem2012synthesis}. (3) Expertise in modular composition and integration, a useful knowledge that system-level functionality and correctness are preserved during module integration.

\subsection{Knowledge Acquisition for Agents}
\label{subsec:knowledge_acquisition}

The effectiveness of the proposed framework depends heavily on the quality of the knowledge injected in each agent. It is essential to endow the agents with deep and broad background knowledge in order to fully leverage their strong reasoning and generation capabilities.

\subsubsection{Real-World Software Projects}

Real-world software projects serve as the richest and most pragmatic source of knowledge for agents within the framework. Among all artifacts in a software system, source code is typically the most complete and well-maintained. We are confident that LLMs at the heart of agents have already internalized diverse programming paradigms, coding idioms, and implementation strategies used across different domains and contexts.

In contrast, NLRs in such projects are often sparse, outdated, or entirely missing. We can reverse-engineer requirements from the code itself, including identifying implicit functional behavior, reconstructing user stories, and deducing system-level constraints based on program logic and structure \cite{sun2024source}. Similarly, architectural model is also frequently absent. Therefore, we can analyze module boundaries, call hierarchies, and data flow patterns to infer the underlying software architecture \cite{2025-MAAD}. Furthermore, informal and formal specifications are rarely present in real-world codebases. As a result, a feasible way is to extract informal specifications by summarizing source code and comments \cite{sun2024source}, and even attempt to derive formal specifications by identifying verifiable properties, preconditions, and invariants directly from implementations \cite{wen2024enchanting,ma2024specgen}. By engaging in the above process, agents can gradually develop a comprehensive understanding of how requirements, architecture, specifications, and code are interrelated. They learn to construct internal mappings among these artifacts and accumulate experience in writing formal specifications that reflect real-world software logic.

\subsubsection{Authoritative Literature}

Authoritative literature forms a foundational pillar of knowledge for agents. These resources provide the theoretical grounding, methodological rigor, and historical context that complement the practical patterns observed in source code.

Classic textbooks and reference works in software engineering offer systematic knowledge across core areas such as requirements engineering, software architecture, coding practices, testing, and project management. These materials define the best practices and design principles that help agents reason about software quality, scalability, and maintainability.

Equally important are texts that bridge formal methods with software engineering. Agents should learn the methodologies for specifying, modeling, and verifying key properties using mathematical rigor. Topics such as model checking \cite{merz2008specification}, deductive verification \cite{acsl_reference}, theorem proving \cite{bertot2013interactive}, abstract interpretation \cite{cousot2021principles}, and SMT (Satisfiability Modulo Theories) \cite{barrett2018satisfiability} are essential to enable the Formalizer and Coder to produce reliable, provably correct artifacts.

\subsubsection{Professional Experts}

Obviously, two major categories of experts are relevant to \ours{}, i.e., software engineers and formal methods experts. Software engineers contribute by validating functional requirements, refining architecture designs, and promoting best practices in code generation. Their experience provides grounded insight into practical development processes, common pitfalls, and robust design patterns. 
Formal methods experts, on the other hand, play a crucial role in verifying specification quality, guiding the formulation of formal properties, and applying proof techniques. They possess the know-how to select and use formal verification tools \cite{cavada2014nuxmv,frama_c_user_manual,beyer2011cpachecker,de2008z3} effectively, interpret verification outcomes, and ensure that specifications are both expressive and provable. By simulating the knowledge and feedback from these experts (e.g., Fine-tuning \cite{hu2022lora}, CoT \cite{wei2022chain} and RAG \cite{lewis2020retrieval}), agents can operate with greater accuracy and adaptability, marrying theoretical rigor with practical feasibility.

\begin{figure*}[t]
    \centering
    \includegraphics[width=1\linewidth]{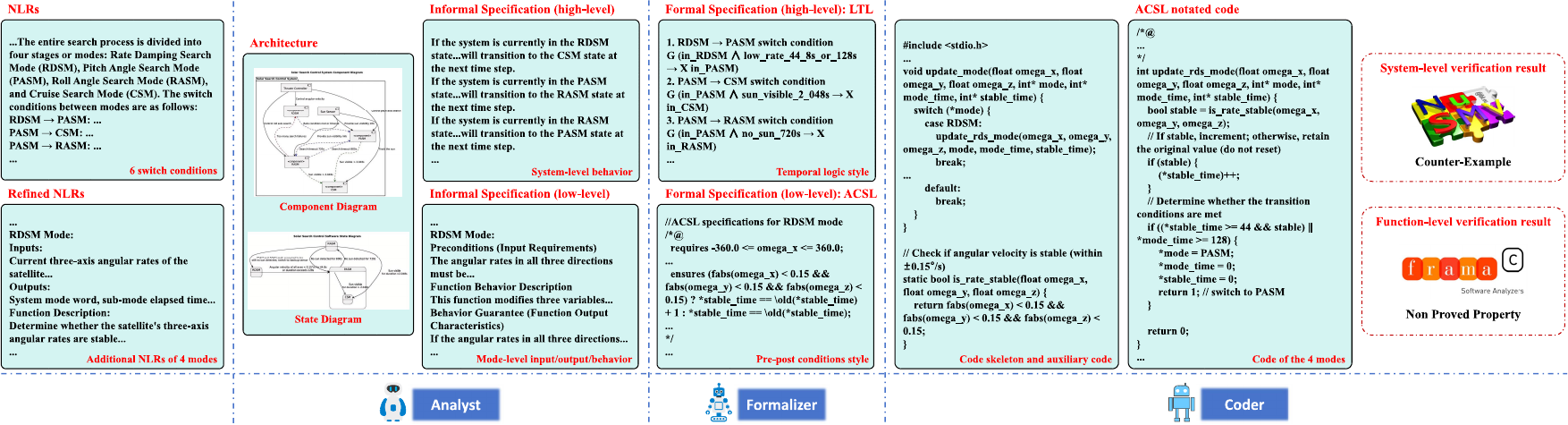}
    \caption{Main artifacts of \ours{} on a practical example}
    \label{fig: case study}
\end{figure*}

\section{Case Study}
\label{sec:case_study}
In this section, we demonstrate the potential capabilities of \ours{} through a practical example. The selected case is typical control software in the safety-critical aerospace domain, which imposes strict constraints on functional correctness, compliance with domain standards, and hardware-dependent operations. The system is named Solar Search (SS), which aims to accomplish the functionality of solar acquisition. SS is composed of several subsystems, with the mode-switching component as the central one. The process of solar acquisition involves four separate stages, namely modes. The four modes switch from each other based on specific conditions. \ours{} is applied to generate the mode-switching component with intermediate outputs to show the stepwise progression. All the artifacts produced through the pipeline are presented in Figure~\ref{fig: case study}.

The coarse NLRs clearly define four modes, along with switch conditions between them. The refined NLRs are built upon the coarse ones, enriched with more detailed inputs, outputs and functional descriptions of the four modes. Upon reviewing the refined NLRs, the Analyst produces a component diagram and a state diagram to represent the architecture. The content extracted from the system-related NLRs is transformed into system-level properties, capturing the high-level behavioral expectations of the system. The content extracted from the mode-specific NLRs is mapped to mode-level properties, focusing on the low-level input, output and behavior within individual operational modes.

The semantic meaning of the natural language is further interpreted by the Formalizer, and necessary constraints are applied to eliminate any remaining ambiguities. Here the high-level informal specifications are transformed into LTL (Linear Temporal Logic) specifications \cite{pnueli1977temporal}, and the low-level informal specifications into ACSL (ANSI/ISO C Specification Language) specifications \cite{baudin2021acsl}.

The Coder generates C code for each mode according to the low-level formal specifications and the refined NLRs. The Coder also generates a code skeleton to integrate the four modes. Subsequently, the Coder performs the verification on two aspects of the system. First, model checking the behavior model of the system (state diagram\footnote{LLMs can easily convert the state diagram into a model representation that is compatible with many formal tools.}) against the LTL specifications using the model checker NuSMV \cite{cimatti2002nusmv}. Second, deductive reasoning is applied to each mode function annotated with ACSL specifications using the program analyzer Frama-C \cite{kirchner2015frama}.

In our experience, humans can be involved in the process of requirement refinement, architecture model verification, and mode-related function verification. Humans assist in revising the NLRs obtained during requirement refinement, based on the user intent elaborated by the Analyst. Architecture model verification reveals inconsistencies between the architecture model and the requirements, which humans can resolve. Mode-related function verification can leverage verification results, with human assistance, to adjust the low-level formal specifications.

\section{Challenges and Opportunities}

\subsection{Completeness and Consistency of Formal Specifications}

One of the key challenges is how to generate formal specifications that are sufficiently complete. What we mean by ``complete" is that formal specifications are capable of specifying all essential aspects of the NLRs. Incomplete specifications may lead to a divergence between what the system is expected to do and what the verification process can guarantee. In practice, NLRs are often ambiguous, under-specified, or inconsistent. In this case, the Formalizer encounters difficulties in inferring all relevant system properties, especially those that are implicit or only loosely defined. Therefore, specification completeness requires not only the ability to parse and formalize explicit requirements but also to infer hidden assumptions, identify missing properties, and reason about potential corner cases.

Another critical issue lies in maintaining consistency between the informal requirements and the corresponding formal specifications. Misalignment may occur due to semantic drift during translation, misunderstanding of domain-specific language, or the inherent difficulty of preserving intent across different levels of abstraction. Such inconsistency undermines the trustworthiness of the entire pipeline. Even if the code is formally verified against the specification, any mismatch between the specification and the original intent still renders the system incorrect from the user’s perspective.

\subsection{High-Quality Code from Formal Specifications}

The challenge pertains to the Coder’s ability to synthesize high-quality, verifiable code from formal specifications. Although formal specifications provide a clear semantic target, the actual process of generating robust, maintainable, and idiomatic code that meets this target is non-trivial. Similar to informal requirements, the semantic gap between formal specifications and executable code still poses a fundamental difficulty. Some specifications may admit many correct implementations, and choosing the most suitable one requires understanding performance constraints, usability factors, and deployment contexts. Moreover, verification tools vary in capability and may impose restrictions on how code must be structured to be verifiable.

\subsection{Practical Deployment}

The real-world deployment of our framework raises several practical challenges. As mentioned earlier, the availability of high-quality training data for each agent remains a concern. Informal and formal specifications are often missing or under-documented in real-world software projects, making it difficult to pretrain or fine-tune the agents effectively. The human-in-the-loop review process, although beneficial for quality assurance, introduces additional complexity. It is non-trivial to define when and how expert feedback should be requested, how it is incorporated into agent decisions, and how to balance automation with human judgment. Scalability and performance pose important engineering concerns. As the system scales to larger projects with numerous components and deeper formal properties, the computational cost of verification and the complexity of code synthesis may grow substantially.

\section{Conclusion}

This paper proposes \ours{} that combines software engineering and formal methods to achieve verifiable ACG. By dividing responsibilities among Analyst, Formalizer, and Coder agents, the framework creates a structured pipeline from NLRs to verified executable code, supporting consistency, traceability, and correctness throughout the software development lifecycle. Aiming to improve each agent’s powerful reasoning and generation capabilities, we emphasize the significance of deep background knowledge, drawn from real-world projects, authoritative literature, and professional experts. While promising, the framework still faces key challenges such as ensuring specification completeness, maintaining semantic alignment, and producing requirement-compliant verified code. Addressing these challenges presents both a research opportunity and a practical path toward more intelligent and trustworthy development of software systems.

\bibliographystyle{IEEEtran}
\bibliography{references}
\balance

\end{document}